\documentstyle[preprint,aps,prd]{revtex}

\begin{document}
\draft 
\title{Thermal partition function of photons and \\
gravitons in a Rindler wedge }
\author{Devis Iellici\footnote{Electronic address: \sl
iellici@science.unitn.it}
  and  Valter Moretti\footnote{Electronic address: 
\sl moretti@science.unitn.it}}
\address{Dipartimento di Fisica, Universit\`a di Trento, \protect\\
and Istituto Nazionale di Fisica Nucleare,\protect\\
Gruppo Collegato di Trento \protect\\ I-38050 Povo (TN), Italy}

\date{July 1996}
\preprint{Preprint UTF - 378}

\maketitle

\begin{abstract}

{\small 
The thermal partition function of photons in any covariant gauge and
gravitons in the harmonic gauge, propagating in a Rindler wedge, are
computed using a local $\zeta$-function regularization approach. The
correct Planckian leading order temperature dependence $T^4$ is
obtained in both cases. For the photons, the existence of a surface
term giving a negative contribution to the entropy is confirmed, as
earlier obtained by Kabat, but this term is  shown to be gauge
dependent in the four-dimensional case and, therefore is  discarded.
It is argued that similar terms could appear dealing with any integer
spin $s\geq 1$ in the massless case and in more general manifolds. Our
conjecture is checked in the case of a graviton in the harmonic gauge,
where  different surface terms also appear, and physically consistent
results arise dropping these terms. The results are discussed in
relation to the quantum corrections to the black hole entropy.}

\end{abstract}


\narrowtext

\section{Introduction}

In recent years, many papers have been concerned with the first
quantum correction to the Bekenstein-Hawking black hole entropy.
According to 't Hooft \cite{TH}, the main contribution to these
corrections comes from quantum fields propagating in the region
outside the horizon. An important tool used to compute these
corrections is the approximation of the metric of a large mass
Schwarzschild black hole given by the simpler Rindler metric. In this
approximation the quantum corrections are identified with the entropy
of thermal states of quantum fields in the Rindler space-time. Many
different methods have been employed to compute this entropy and,
among them, the method of the conical singularity is one of the most
used: one follows the usual prescription to compute the thermal
partition function of a quantum field, that is to evaluate the
Euclidean path integral over all the field configurations that are
periodic in the imaginary time and identify the period $\beta$ with
the inverse of the temperature. In doing this, the Rindler manifold
acquires a conical singularity with angular deficit $2\pi-\beta$, and
so one sees that, in order to avoid the singularity, there is only one
possible temperature for the system, i.e. the Unruh-Hawking
temperature $\beta=2\pi$. However, if one wants to compute
thermodynamical quantities such as the entropy and the internal energy
using standard thermodynamical relations such as $
S_\beta=\beta^2\partial_\beta F_\beta$, then one needs to go ``off
shell'', i.e., consider $\beta\neq2\pi$ and so manifolds with a
conical singularities. Therefore, many techniques have been developed
to compute the one-loop quantum corrections on manifolds with conical
singularity. In this respect, it is important to note that the
standard use of heat-kernel plus proper-time regularization yields the
wrong temperature dependence of the free energy and the other
thermodynamical quantities, at least when the dimension of the
space-time is not two \cite{RE}. In four dimensions, in particular,
the leading term in the high temperature limit of the free energy
should be Planckian, namely, proportional to $\beta^{-4}$
\cite{DO78,DO94,SU}, while the heat kernel gives $\beta^{-2}$
independently of the dimension.

In this context, Zerbini, Cognola and Vanzo \cite{ZCV}, starting from
a previous work of Cheeger \cite{JC}, have recently introduced a new
method to compute the effective action of a scalar field on manifolds
with conical singularities using the $\zeta$-function regularization.
This method, in addition to  giving the correct temperature dependence
and allowing one to work directly with massless fields, has the
advantage that it does not require the regularization of the conical
singularity or transforming the cone in a compact manifold, procedures
which do not have a clear physical meaning if one is interested in the
(Euclidean) Rindler space. The drawbacks are that this method is
technically difficult to apply in the case of massive fields and
especially that it yields for the part of the free energy proportional
to $\beta^{-2}$ a numerical coefficient different from that obtained
with the point splitting and the optical metric methods
\cite{DO78,BO,FS,ALO}. This latter problem is shared with the
heat-kernel approach and the reason for this discrepancy is not yet
understood.

Most of the work on the quantum corrections to the black hole entropy
is carried on using the scalar field. Results for higher spins have
been obtained translating earlier results obtained for the closely
related cosmic string background \cite{DO94}. Last year, in an
interesting paper \cite{DK} Kabat investigated the corrections
to the black hole entropy coming form scalar, spinor and vector fields
by explicitly writing the field modes in the Euclidean Rindler space
and then using the heat-kernel and the proper-time regularization. In
the vector field case  he has obtained an unexpected ``surface''
term, which corresponds to particle paths beginning and ending at the
horizon. This term gives a negative contribution to the entropy of the
system and, in fact, is large enough to make the total entropy
negative at the equilibrium temperature. Kabat argues that this term
corresponds to the low-energy limit of string  processes which couple
open strings with both ends attached to the horizon and closed strings
propagating outside the horizon diagrams and discussed by Susskind and
Uglum \cite{SU} as responsible for black hole entropy within string
theory.

 In this paper, we apply the method of \cite{ZCV} to the case of the
Maxwell field and the graviton field. As a result, in the case of the
photon field we confirm that  there is a `surface term' which would
give a negative contribution to the entropy, as obtained by Kabat in
\cite{DK}. However, beside getting a different temperature dependence,
we show that it depends on the gauge-fixing parameter and so we
discuss how it is possible to discard it. In this way we also avoid
embarrassing negative entropies. In the case of the graviton we get
similar surface terms and show that one can get consistent physical
results by discarding them. We also discuss the appearance of similar
terms in more general manifolds. After discarding the surface terms we
get the reasonable result that the effective action and all the
thermodynamical quantities are just twice those of the minimally
coupled scalar field: this is in agreement with the results of the
point-splitting method \cite{FS,ALO}, the heat kernel method
\cite{LW,BS,FM}, and, apart from the surface terms, also with Kabat
\cite{DK}.

We remind that the Rindler wedge is  a globally hyperbolic manifold
defined by the inequality $x>|t|$, in the usual set of rectangular
coordinates $(t,x,y,z)$ of Minkowski space-time. In this wedge we can
define a new set of static coordinates by setting $t=r\sinh\tau$ and
$x=r\cosh\tau$, with $0<r<\infty$ and $-\infty<\tau<\infty$. Then the
Minkowski metric takes the form of the Rindler metric:
\begin{eqnarray}ds^2=-r^2 d\tau^2+dr^2+dy^2+d^2z.
\label{Rindler}
\end{eqnarray}
One can see that lines of constants $r,y$, and $z$ are trajectories of
uniformly accelerated particles, with proper acceleration $a=r^{-1}$.

As we said above, the  importance of the Rindler metric is mainly due
to the fact that it can be seen as an approximation of the metric of a
large mass Schwarzschild black hole outside the event horizon. Indeed,
consider the Schwarzschild metric, which describes an uncharged,
nonrotating black hole of mass $M$
\begin{eqnarray}
ds^2=-\left(1-\frac{2GM}{R}\right ) dT^2+\left(1-\frac{2GM}{R}\right
)^{-1} dR^2+R^2 d\Omega_2, \hspace{1cm}
d\Omega_2=d\theta^2+\sin\theta\,
d\varphi^2, \nonumber
\end{eqnarray} 
where $M$ is the mass of the black hole. In the region outside the
event horizon, namely, ${2GM<R<\infty}$, we can define new
coordinates $\tau$ and $r$ by
\begin{eqnarray}
\tau&=&\frac{T}{4GM},\\
r&=&\sqrt{8GM(R-2GM)},
\end{eqnarray}
and so the metric takes the form
\begin{eqnarray}
ds^2&=&-r^2\left(1+\frac{r^2}{16G^2 M^2}\right )^{-1}d\tau^2+
\left(1+\frac{r^2}{16G^2 M^2}\right )dr^2\\
& &+4G^2 M^2\left(1+\frac{r^2}{16G^2 M^2}\right )^2d\Omega_2.
\end{eqnarray}
If we take the large mass limit, the last term becomes the metric of a
spherical surface with very large radius that can be approximated by a
flat metric $dy^2+dz^2$. Then, in this limit, the metric becomes the
Rindler one, Eq. (\ref{Rindler}). Actually, even if we do not
consider the large mass limit, the approximation should become better
and better as we approach the event horizon, $r=0$.

The Rindler metric is also related with the study of the cosmic string
background: the metric around an infinitely long, static, straight and
with zero thickness cosmic string can be written as
\begin{eqnarray}
ds^2=-dt^2+dz^2+dr^2+r^2d\varphi,\hspace{1cm}0\leq\varphi\leq\alpha, 
\nonumber
\end{eqnarray}
where the polar angle deficit $2\pi-\alpha$ is related to the mass per
unit length of string $\mu$ by $2\pi-\alpha=8\pi G\mu$. Since the
metric is ultrastatic, we can perform a Wick rotation, $t \rightarrow
it$, and the metric becomes equal to the Euclidean Rindler metric.
Therefore, we can identify the thermal partition function of a field
at temperature $\alpha^{-1}$ in the Rindler wedge with the
zero-temperature Euclidean-generating functional of the same field in
a cosmic string background.

The rest of this paper is organized as follows. In Sec. II we
compute the one-loop effective action for the electromagnetic field on
the manifold $C_\beta\times\mbox{R}^2$ using the $\zeta$-function
regularization. We use this result to compute the quantum correction
to the black hole entropy in the framework of conical singularity
method. In Sec. III we formulate a general conjecture on the
appearance of Kabat-like surface terms in the case of integer spin and
general manifolds. In Sec. IV the conjecture is checked in the case
of the graviton. Sec. V is devoted to the discussion of the
results.

\section{EFFECTIVE ACTION FOR THE PHOTON FIELD}

In a curved space-time with Lorentz signature the action of the
electromagnetic field is  $S=\int{\cal L}(x) \sqrt{-g}d^4x$, 
where the Lagrangian scalar density\footnote{We
adopt the convention that the indices $a,b,\dots=\tau,r,y,z$ are for
the whole manifold, the greek indices are for the pure cone,
$a,b,\dots=\tau,r$, and the indices $i,j,\dots=y,z$ are for the
transverse flat directions.} is \cite{BD}
\begin{eqnarray}
{\cal L}_{\mbox{\scriptsize em}}(x)&=&-\frac{1}{4}
F_{ab}F^{ab},\nonumber\\
F_{ab}&=&\nabla_a A_b-\nabla_b A_a=
\partial_aA_b-\partial_a A_b.
\label{em1}
\end{eqnarray}
We need also the gauge-fixing term and the contribution of the
ghosts:
\begin{eqnarray}
{\cal L}_G&=&-\frac{1}{2\alpha}(\nabla^a A_a)^2,\\
{\cal L}_{\mbox{\scriptsize ghost}}&=&\frac{1}{\sqrt{\alpha}}
g^{ab}\partial_a c\partial_b c^{\ast},
\label{em2}
\end{eqnarray}
where $c$ and $c^\ast$ are anticommuting scalar fields. The dependance
on the gauge-fixing parameter $\alpha$ is relevant only in presence of 
a scale anomaly. SInce this is not the case here, we shall ignore it in the
rest of this paper.

We are interested  in the finite temperature theory and so we change
$\tau\rightarrow
i\tau$ and identify $\tau$ and $\tau+\beta$. The metric of the
Rindler space-time turns to Euclidean signature,
$ds^2=r^2d\tau^2+dr^2+dy^2+dz^2$, and the vector D'Alembertian
operator $\Box$ becomes the vector Laplace-Beltrami operator $\Delta$.
In the following this operator will be simply called Laplacian. The
one-loop effective action for this theory will then be given by the
following determinants:
\begin{eqnarray}
\ln Z_\beta=-\frac{1}{2}\ln\det \mu^{-2}\left (g^{ab}(-\Delta)-R^{ab}+
(1-\frac{1}{\alpha})\nabla^{a}\nabla^{b}\right )+\ln
Z_{\beta,\mbox{\scriptsize ghosts}},
\label{em10}
\end{eqnarray}
where $\mu^2$ is the renormalization scale and the effective action of
the ghosts is minus twice the effective action of a scalar massless
field, which is well known \cite{JC}\cite{ZCV}. It is important to note
that the determinant has to be evaluated on the whole set of
eigenfunctions, not only on the physical ones \cite{DV}.

We work on the manifold $C_\beta\times \mbox{R}^2$,
where $C_\beta$ is
the cone with angular deficit equal to $2\pi-\beta$. This manifold is
flat everywhere but on the tip of the cone, where the curvature has a
$\delta$-function singularity. Nevertheless,  the modes we use vanish
on the tip, and so we can consider $R_{ab}=0$. Note also that, due to
the flatness, the covariant derivatives commute. Hence, we  are left
with the problem of computing the determinant of the operator
$[g^{ab}(-\Delta)+ (1-\frac{1}{\alpha})\nabla^{a}\nabla^{b}]$ acting
on vectors. In order to define this determinant we use the
$\zeta$-function regularization: first, suppose we have a complete set
of eigenfunctions of the operator, indicated as $A_a^{(i,n\lambda{\bf
k })}(x)$, with eigenvalue $\nu_i^2(n\lambda{\bf k})$. Here, ${\bf
k}=(k_y,k_z)$, $a=\tau,r,y,z$ and $i=1,\dots,4$ is the polarization
index. In this notation we have taken into account the triviality of
the transverse dimension and the fact that we have a discrete index
$n$ since the $\tau$ coordinate is compact and we impose periodic
boundary conditions. Then we can define the local, diagonal heat
kernel as
\begin{eqnarray}
K^{(i)}(t;x)=\sum_n\int d\mu(\lambda)\,d^2{\bf k}\;
e^{-t\nu_i^2} g^{ab} A_a^{(i)}(x)A_b^{(i)\ast}(x),
\label{em12}
\end{eqnarray}
where $d\mu(\lambda)$ is an appropriate integration measure. The
corresponding local spin-traced $\zeta$ function can be obtained
through a Mellin transform:
\begin{eqnarray}
\zeta(s;x)=\frac{1}{\Gamma(s)}\int_0^\infty dt\, t^{s-1}
\sum_i K^{(i)}(t;x).
\label{em16}
\end{eqnarray}
Alternatively, we can define the local $\zeta$ function as the inverse
power of the kernel of the above differential operator: the spectral
representation gives directly
\begin{eqnarray}
\zeta(s;x)=\sum_i\sum_n\int d\mu(\lambda)\,d^2{\bf k}\;
[\nu_i^2(n\lambda{\bf k})]^{-s} g^{ab} A_a^{(i)}(x)A_b^{(i)\ast}(x)
\label{emZETA}
\end{eqnarray}
In general, both the Mellin transform and the inverse power of the
operator require analytic continuation arguments to be defined at the
physical values of $s$.

We can also define a global $\zeta$ function by tracing over the space
indices:
\begin{eqnarray}
\zeta(s)= \int d^4x \sqrt{g} \, \zeta(s;x).
\label{em17}
\end{eqnarray}
This last step is delicate: in general, the operation of tracing over
the space indices requires the introduction of a smearing function,
since the manifold is noncompact and there can be nonintegrable
singularities in the local $\zeta$ function, and a particular choice of
the smearing function could sweep away important information. This is
one of the reasons why we prefer to work with a local formalism as
long as possible. Once we have computed and analytically continued the
$\zeta$ function, we can write the effective lagrangian density and the
effective action as
\begin{eqnarray}
{\cal L}_\beta(x)&=&\frac{1}{2}
\zeta'(s=0;x)+\frac{1}{2}\zeta(s=0;x)\ln \mu^2,\nonumber\\
\ln Z_{\beta}&=&\int d^4 x \sqrt{g}\,{\cal L}_\beta(x).
\label{em18}
\end{eqnarray}
Of course, to the above expression we have to add the contribution of
the ghosts, which is minus two times the effective lagrangian density
of a scalar field.

A suitable set of normalized eigenfunctions of the operator
$[g^{ab}(-\Delta)+(1-\frac{1}{\alpha})\nabla^a\nabla^b]$ (equivalent to
Kabat's set \cite{DK} if $\alpha=1$) is the following: setting
$k=|{\bf k}|$
\begin{eqnarray}
A_a^{(I,n\lambda{\bf k})}&=&\frac{1}{k}\epsilon_{ij}\partial^j\phi=
\frac{1}{k}(0,0, ik_z\phi,-ik_y\phi)\nonumber,\\
A_a^{(II,n\lambda{\bf
k})}&=&\frac{\sqrt{g}}{\lambda}\epsilon_{\mu\nu}\nabla^\nu\phi=
\frac{1}{\lambda}(r\partial_r\phi,-\frac{1}{r}\partial_\tau\phi,0,0),
\nonumber\\
A_a^{(III,n\lambda{\bf k})}&=&\frac{1}{\sqrt{\lambda^2+{\bf k}^2}}
(\frac{k}{\lambda}\nabla_\mu-\frac{\lambda}{k}\partial_i)\phi=
\frac{1}{\sqrt{\lambda^2+{\bf k}^2}}
(\frac{k}{\lambda}\partial_\tau\phi,\frac{k}{\lambda}\partial_r\phi,
-\frac{\lambda}{k}\partial_y\phi,-\frac{\lambda}{k}\partial_z\phi),
\nonumber\\
A_a^{(IV,n\lambda{\bf k})}&=&\frac{1}{\sqrt{\lambda^2+{\bf k}^2}}
\nabla_a\phi=\frac{1}{\sqrt{\lambda^2+{\bf k}^2}}
(\partial_\tau\phi,\partial_r\phi,\partial_y\phi,\partial_z\phi),
\label{em20b}
\end{eqnarray}
where $\sqrt{g}\epsilon_{\mu\nu}$ is the Levi-Civita pseudotensor on
the cone, $\epsilon_{ij}$ is the Levi-Civita pseudo-tensor on R$^2$ in
Cartesian coordinates, and $\phi=\phi_{n\lambda{\bf k}}(x)$ is the
complete set of normalized eigenfunctions of the  Friedrichs
self-adjoint extension of the scalar Laplacian on $C_\beta\times
\mbox{R}^2$
\cite{KS}:
\begin{eqnarray}
\phi_{n\lambda{\bf k}}(x)&=&\frac{1}{2\pi\sqrt{\beta}}e^{ik_y y+ik_z z}
e^{i\frac{2\pi n}{\beta}\tau} J_{\nu_n}(\lambda r),\hspace{5mm}
n=0,\pm 1, \dots;\,\,\lambda\in \mbox{R}^+;\,\, k_y, k_z\in \mbox{R}
\nonumber\\
\Delta\phi_{n\lambda{\bf k}}(x)&=&-(\lambda^2+{\bf
k}^2)\phi_{n\lambda{\bf k}}(x).
\label{em21}
\end{eqnarray}
Here, $J_{\nu_n}$ is the Bessel function of first kind and
$\nu_n=\frac{2\pi|n|}{\beta}$. Using the relation
\begin{eqnarray}
\int_0^\infty dr\,r\, J_\nu(\lambda' r)J_\nu(\lambda
r)=\frac{1}{\lambda}\delta(\lambda-\lambda'), \nonumber
\end{eqnarray}
one can check that the  modes (\ref{em20b}) are normalized according
to
\begin{eqnarray}
(A^{(i',n'\lambda'{\bf k}')},A^{(i,n\lambda{\bf k})})&\equiv&\int
d^4x\,\sqrt{g}\,g^{ab} A_a^{(i',n'\lambda'{\bf
k}')\ast}A_b^{(i,n\lambda{\bf k})}\nonumber \\
&=&\delta_{i'i}\delta_{n'n}\delta^{(2)}({\bf k}-{\bf
k}')\frac{1}{\lambda}\delta(\lambda-\lambda'),\nonumber
\end{eqnarray}
The first three eigenfunctions (\ref{em20b}) satisfy $\nabla^a A_a=0$
and have eigenvalue $\lambda^2+{\bf k}^2$, while $A_a^{(IV)}$ is a pure
gauge and has eigenvalue $\frac{1}{\alpha}(\lambda^2+{\bf k}^2)$.

Using these eigenfunctions, we can compute the diagonal
$\zeta$ function using the spectral representation Eq. (\ref{emZETA}):
after the integration over $d{\bf k}$, the contributions of the modes
to the diagonal $\zeta$ function are
\begin{eqnarray}
\zeta^{(I)}(s;x)&=&\zeta^{\mbox{\scriptsize scalar}}(s;x),\nonumber\\
\zeta^{(II)}(s;x)&=&\frac{\Gamma(s-1)}{4\pi\beta\Gamma(s)}\sum_n
\int_0^\infty
d\lambda\,\lambda^{1-2s}[\frac{\nu_n^2}{r^2}J_{\nu_n}^2(\lambda
r)+(\partial_r J_{\nu_n}(\lambda r))^2],\nonumber\\
\zeta^{(III)}(s;x)&=&\!\!\frac{s-1}{s}\zeta^{\mbox{\scriptsize
scalar}}(s;x)+
\frac{\Gamma(s-1)}{4\pi\beta\Gamma(s+1)}\sum_n\int_0^\infty
d\lambda\lambda^{1-2s}[\frac{\nu_n^2}{r^2}J_{\nu_n}^2(\lambda
r)+(\partial_r J_{\nu_n}(\lambda
r))^2],\nonumber\\
\zeta^{(IV)}(s;x)&=&\frac{\alpha^s}{s}\zeta^{\mbox{\scriptsize
scalar}}(s;x)+
\frac{\alpha^s\Gamma(s)}{4\pi\beta\Gamma(s+1)}\sum_n\int_0^\infty
d\lambda\,\lambda^{1-2s}[\frac{\nu_n^2}{r^2}J_{\nu_n}^2(\lambda
r)+(\partial_r J_{\nu_n}(\lambda r))^2],\nonumber
\end{eqnarray}
where the spectral representation of the local $\zeta$ function of a
minimally coupled scalar field on $C_\beta\times \mbox{R}^2$ is
\begin{eqnarray}
\zeta^{\mbox{\scriptsize scalar}}(s;x)=
\frac{\Gamma(s-1)}{4\pi\beta\Gamma(s)}\sum_{n=-\infty}^{\infty}
\int_0^\infty
d\lambda\,\lambda^{3-2s} J_{\nu_n}^2(\lambda r).\nonumber
\end{eqnarray}
Now, looking for a way close to that followed by Kabat \cite{DK},
we use the following identity, which can be proved using some
recursion formulas for the Bessel functions \cite{GR},
\begin{eqnarray}
2\left [ \frac{\nu_n^2}{r^2}J_{\nu_n}^2(\lambda r)+[\partial_r
J_{\nu_n}(\lambda r)]^2 \right ]= 2\lambda^2 J_{\nu_n}^2(\lambda
r)+\frac{1}{r}\partial_r r\partial_r J_{\nu_n}^2(\lambda r),
\label{em25}
\end{eqnarray}
and so the spin-traced local $\zeta$ function becomes
\begin{eqnarray}
\zeta(s;x)&=&(1+\frac{s-1}{s}+\frac{\alpha^s}{s})
\zeta^{\mbox{\scriptsize
scalar}}(s;x)  \nonumber \\
& &+\frac{s+1+\alpha^s(s-1)}{2s}
\frac{\Gamma(s-1)}{4\pi\beta\Gamma(s)}\sum_n
\int_0^\infty \,\lambda^{1-2s}[2\lambda^2 J_{\nu_n}^2(\lambda r)+
\frac{1}{r}\partial_r
r\partial_r J_{\nu_n}^2(\lambda r)] \:,\nonumber
\end{eqnarray}
namely
\begin{eqnarray}
\zeta(s;x) &= &(3+\alpha^s)\zeta^{\mbox{\scriptsize scalar}}(s;x)+
\frac{s+1+\alpha^s(s-1)}{2s}\zeta^{\mbox{\scriptsize V}}(s;x), 
\label{ema0}
\end{eqnarray}
where we have set
\begin{eqnarray}
\zeta^{\mbox{\scriptsize V}}(s;x)=\frac{1}{r}\partial_r r\partial_r
\frac{\Gamma(s-1)}{4\pi\beta\Gamma(s)}\sum_{n=-\infty}^{\infty}
\int_0^\infty d\lambda\,\lambda^{1-2s}J_{\nu_n}(\lambda r)^2.
\label{em101}
\end{eqnarray}
Notice that the term $\zeta^{\mbox{\scriptsize V}}(s;x)$ arises from
the ``conical'' components of the field, i.e. $A_\tau$ and $A_r$. In
particular its source is the second term in the right-hand side of Eq.
(\ref{em25}) only. This terms will produce the Kabat ``surface term''
as we will see shortly.

We have taken $\frac{1}{r}\partial_r r\partial_r$, which is in fact
the Laplacian $\Delta$, outside the integral and the series, but this
is a safe shortcut: indeed, one could first let $\Delta$ act on the
Bessel function using $\partial_r J_\nu(\lambda r)=\lambda
J_{\nu-1}(\lambda r)-\frac{\nu}{r}J_\nu(\lambda r)$, go through some
tedious calculations and get the same result as Eq. (\ref{em27}).

So far, the expressions for $\zeta^{\mbox{\scriptsize scalar}}$ and
$\zeta^{\mbox{\scriptsize V}}$ are just formal, since one can easily
see that there is no value of $s$ for which they converge.  The
correct way to compute $\zeta^{\mbox{\scriptsize scalar}}$ in this
background has been recently given by Zerbini, Cognola and Vanzo
\cite{ZCV}, following an earlier work of Cheeger \cite{JC}, and the
result is
\begin{eqnarray}
\zeta^{\mbox{\scriptsize scalar}}(s;x)=
\frac{r^{2s-4}}{4\pi\beta\Gamma(s)}
I_\beta(s-1), \nonumber
\end{eqnarray}
where
\begin{eqnarray}
I_\beta(s)&=&\frac{\Gamma(s-\frac{1}{2})}{\sqrt{\pi}}
[G_\beta(s)-G_{2\pi}(s)],\\
G_\beta(s)&=&\sum_{n=1}^\infty\frac{\Gamma(\nu_n-s+1)}
{\Gamma(\nu_n+s)},
\hspace{1cm} G_{2\pi}(s)=-\frac{\Gamma(1-s)}{2\Gamma(s)},
\nonumber\\
I_\beta(0)&=&\frac{1}{6}\left(\frac{2\pi}{\beta}-
\frac{\beta}{2\pi}\right),\nonumber\\
I_\beta(-1)&=&\frac{1}{90}\left(\frac{2\pi}{\beta}-
\frac{\beta}{2\pi}\right)
\left[\left(\frac{2\pi}{\beta}\right)^2+11\right] \nonumber . 
\end{eqnarray}
The function $I_\beta(s)$ is analytic in the whole complex plane but
in $s=1$, where it has a simple pole with residue
$\frac{1}{2}(\frac{\beta}{2\pi}-1)$. Following the same procedure used
in \cite{ZCV} to obtain the above result, we can compute the
contribution to the $\zeta$ function coming from
$\zeta^{\mbox{\scriptsize V}}(s;x)$. The essential step to give a
sense to Eq. (\ref{em101}) is the separation of the small eigenvalue
$\nu_0$ from the others \cite{JC}: define
\begin{eqnarray}
\zeta_<^{\mbox{\scriptsize
V}}(s;x)&=&\Delta\frac{\Gamma(s-1)}{4\pi\beta \Gamma(s)}
\int_0^\infty
d\lambda\,\lambda^{1-2s}J_0^2(\lambda r), \nonumber \\
\zeta_>^{\mbox{\scriptsize V}}(s;x)&=&2\Delta
\frac{\Gamma(s-1)}{4\pi\beta \Gamma(s)}\sum_{n=1}^{\infty}
\int_0^\infty d\lambda\,\lambda^{1-2s}J_{\nu_n}^2(\lambda r).
\nonumber
\end{eqnarray}
The integrals over $\lambda$ can be computed \cite{GR}: for 
$\frac{1}{2}<\mbox{Re} s<1+\nu$
\begin{eqnarray}
\int_0^\infty  d\lambda\,\lambda^{1-2s} J^2_\nu(\lambda
r)=r^{2s-2}\frac{\Gamma(s-\frac{1}{2}) \Gamma(\nu-s+1)}{2\sqrt{\pi}
\Gamma(s)\Gamma(\nu+s)}. \nonumber
\end{eqnarray}
Therefore, in the strip $\frac{1}{2}<\mbox{Re} s<1$ we get
\begin{eqnarray}
\zeta_<^{\mbox{\scriptsize
V}}(s;x)=-\Delta\frac{r^{2s-2}\Gamma(s-1)}{4\pi\Gamma(s)^2}
\frac{\Gamma(s-\frac{1}{2})}{\sqrt{\pi}}G_{2\pi}(s), \nonumber
\end{eqnarray}
while
\begin{eqnarray}
\zeta_>^{\mbox{\scriptsize
V}}(s;x)=\Delta\frac{r^{2s-2}\Gamma(s-1)}{4\pi\Gamma(s)^2}
\frac{\Gamma(s-\frac{1}{2})}{\sqrt{\pi}}G_{\beta}(s), \nonumber
\end{eqnarray}
which is valid in the strip $1<\mbox{Re} s<1+\nu_1$, since the series
defining $G_\beta(s)$ converges for $s>1$. Both expressions can now be
analytically continued the whole complex plane and then summed, so
we can write
\begin{eqnarray}
\zeta^{\mbox{\scriptsize
V}}(s;x)&=&\Delta\frac{r^{2s-2}\Gamma(s-1)}{4\pi\Gamma(s)^2}
I_\beta(s)\nonumber\\
&=&\frac{(s-1)r^{2s-4}}{\pi\beta\Gamma(s)}I_\beta(s).
\label{em27}
\end{eqnarray}
This result could be obtained directly from Eq. (\ref{em101}), noting
that 
\begin{eqnarray}
\zeta^{\mbox{\scriptsize V}}(s;x)=\Delta[\frac{s}{s-1}
\zeta^{\mbox{\scriptsize scalar}}(s+1;x)]\:.\nonumber
\end{eqnarray}
 Note also that
$\zeta^{\mbox{\scriptsize V}}(s;x)|_{\beta=2\pi}=0$ and
$\zeta^{\mbox{\scriptsize V}}(s=0;x)=0$.

Now we can write the final result for the local $\zeta$ function of the
electromagnetic field: after adding the contribution of the ghosts,
which is just $-2\zeta_\beta^{\mbox{\scriptsize scalar}}(s;x)$, we get
\begin{eqnarray}
\zeta^{\mbox{\scriptsize e.m.}}(s;x)&=&
(1+\alpha^s)\zeta^{\mbox{\scriptsize scalar}}(s;x)+
\frac{s+1+\alpha^s(s-1)}{2s}\zeta^{\mbox{\scriptsize
V}}(s;x)\nonumber\\
&=&(1+\alpha^s)\frac{r^{2s-4}}{4\pi\beta\Gamma(s)}
I_\beta(s-1)+\frac{s+1+\alpha^s(s-1)}{2s}
\frac{(s-1)r^{2s-4}}{\pi\beta\Gamma(s)}I_\beta(s).
\label{em28}
\end{eqnarray}
{From} this expression we can easily see that
$\zeta^{\mbox{\scriptsize e.m.}} (s;x)|_{s=0}=0$ and
\begin{eqnarray}
{\zeta^{\mbox{\scriptsize e.m.}}}'(s;x)|_{s=0}=
\frac{1}{2\pi\beta
r^4}I_\beta(-1)-(1-\frac{1}{2}\ln\alpha)\frac{1}{\pi\beta
r^4}I_\beta(0),
\label{em29}
\end{eqnarray}
Therefore, the one-loop effective Lagrangian density for the
electromagnetic field on $C_\beta\times \mbox{R}^2$ is
\begin{eqnarray}
{\cal L}_\beta^{\mbox{\scriptsize e.m.}}(x)&=&2{\cal
L}_\beta^{\mbox{\scriptsize
scalar}}(x) -\frac{(1-\frac{1}{2}\ln\alpha)}{2\pi\beta
r^4}I_\beta(0)\nonumber\\ &=&\frac{1}{4\pi\beta
r^4}I_\beta(-1)-\frac{(1-\frac{1}{2}\ln\alpha)}{2\pi\beta
r^4}I_\beta(0).
\label{em30}
\end{eqnarray}
Since $I_{2\pi}(s)=0$,  we can notice that both terms of the effective
Lagrangian density vanish when the conical singularity disappears,
$\beta=2\pi$.

A few remarks  on this result. First, no surprise that in in the
effective Lagrangian density we get a contribution which is  twice
that of a scalar field. More surprising is the second term: after the
integration over the spatial variables, it gives rise to what Kabat
\cite{DK} calls ``surface'' term and interprets as a low-energy relic
of stringy effects foreseen by Susskind and Uglum \cite{SU}. This term
would give a negative contribution to the entropy of the system, at
least for  for $\alpha<e^2$, and actually also the total correction to
the entropy at the black hole temperature $\beta=2\pi$ would be
negative for $\alpha<e^{6/5}$, which is clearly a nonsense if we want
to give a state counting interpretation to the entropy. However, in
the four-dimensional case we get that it is not gauge invariant, in
contrast with Kabat's result.

With this regard, it is interesting to note that in two dimensions, i.e.,
on $C_\beta$, the result is indeed independent on the gauge-fixing
parameter: using the modes of the e.m. field on $C_\beta$ given by Kabat
\cite{DK} and following the same procedure as above, before adding the
contribution of ghosts we get
\begin{eqnarray}
\zeta_{d=2}^{\mbox{\scriptsize e.m.}}(s;x)=(1+\alpha^s)
\left [\zeta_{d=2}^{\mbox{\scriptsize scalar}}(s;x)+
\zeta_{d=2}^{\mbox{\scriptsize V}}(s;x)\right ], \nonumber
\end{eqnarray}
where
\begin{eqnarray}
\zeta_{d=2}^{\mbox{\scriptsize
scalar}}(s;x)&=&\frac{r^{2s-2}}{\beta\Gamma(s)} I_\beta(s),\nonumber
\\
\zeta_{d=2}^{\mbox{\scriptsize
V}}(s;x)&=&\Delta\frac{r^{2s}}{2\beta\Gamma(s+1)} I_\beta(s+1),
\nonumber \end{eqnarray}
and so, adding the contribution of the ghosts we have
\begin{eqnarray}
{\cal L}^{\mbox{\scriptsize e.m.}}(x)=\frac{1}{2\pi\beta
r^2}(2\pi-\beta),\nonumber
\end{eqnarray}
which is gauge independent and, after the integration over the
manifold, gives exactly the result of Kabat.

Coming back to the four-dimensional case, we argue that a natural
(albeit not the only possible, see the final discussion)
procedure to restore the gauge invariance is simply to drop the Kabat
term, namely, the last term in Eq. (\ref{em30}), obtaining the
reasonable result ${\cal L}^{\mbox{\scriptsize e.m.}}(x) = 2{\cal
L}^{\mbox{\scriptsize scalar}}(x)$.

First of all, notice that the gauge invariance must  hold for the
integrated quantities as the effective action, namely the logarithm of
the integrated effective Lagrangian. In fact, the ghost procedure,
which takes into account the gauge invariance,  works on integrated
quantities. However, in our case, the integration of the Kabat term
produces a divergent {\em gauge-dependent}  result, and thus it seems
reasonable to discard such a local term. With this regard, it is
important to note that Kabat obtains a gauge-independent result
because, within his regularization procedure, he has the freedom to
choose an independent cutoff parameter for each mode. Instead, in our
procedure we have only one cutoff parameter $\epsilon$, to which we
give a precise physical meaning, namely the minimal distance from the
horizon.

A more general discussion might be the following. It is worth one's
while stressing that, dealing with {\em smooth compact} manifold,
local quantities as local heat kernel and local zeta-functions are
intrinsically ill defined due to the possibility of adding to them a
total covariant derivative with vanishing integral. In such a case,
the previous global quantities are well-defined, and one can
satisfactorily employ these latter instead of local quantities in
order to avoid the ill-definiteness problem.
Notice also that the gauge dependent Kabat surface term formally looks
such as a Laplacian and thus it should disappear after a global
integration, provided regularity conditions on the manifold are
satisfied, producing gauge-independent integrated quantities. However,
this is not the case for the present situation, where the background
is a noncompact manifold with a conical singularity, and the
integrated quantities diverge  requiring a regularization procedure.
We stress that the use of local quantities is preferred on the
physical ground, because they lead us to the correct temperature
dependency as we will see shortly.

Therefore, in our case  the local quantities remain ill defined  and
require a further regularization procedure in order to fix the
possible added total derivative term before we integrate. Furthermore,
the integrated quantities are divergent, so we expect we to have to
take into account also total derivative terms with a divergent
integral.  In our case this further regularization procedure consists
just in discarding the Kabat term. Notice that this procedure produces
gauge-independent local quantities.

Once we have dropped the Kabat's term, we can compute thermodynamical
quantities like internal energy and entropy: we need the effective
action and so we have to introduce a smearing function $\varphi(x)$ in
order to define the trace: $\ln Z_\beta=\int d^4x \sqrt{g}{\cal
L}_\beta(x)\varphi(x)$. Actually, since ${\cal L}_\beta$ does not
depend on the transverse coordinates $y$ and $z$, the integration on
these coordinates simply yields the infinite area of the Rindler
horizon, that we indicate as $A_\perp$. This divergence has clear
physical meaning. The integration over $\tau$ has no problem, while a
convenient smearing function for  the integration over $r$ is
$\varphi(r)=\theta(r-\epsilon)$, and so the effective action becomes
\begin{eqnarray}
\ln Z_\beta(\epsilon)=\frac{A_\perp}{8\pi\epsilon^2}I_\beta(-1).
\label{em31}
\end{eqnarray}
For $\epsilon \rightarrow 0$ we have a divergence that can be seen as
a ``horizon'' divergence \cite{TH}, since as $r\rightarrow 0$ we
approach the horizon of the Rindler wedge.

{From} Eq. (\ref{em31}) we can compute the free energy,
$F_\beta=-\frac{1}{\beta}\ln Z_\beta$, which at high temperature,
$\beta\rightarrow 0$, has a leading behavior
$-2\times\frac{\pi^2A_\perp}{180\epsilon^2\beta^4}$, in perfect
agreement with the statistical mechanics result of  Susskind and Uglum
\cite{SU}. Instead, Kabat \cite{DK} obtains a leading behavior
$-2\times\frac{A_\perp}{8\epsilon^2\beta^2}$, where the behavior
$\beta^{-2}$ independent of the dimension of the space-time, is
typical of the integrated heat-kernel approach. Of particular interest
for the black hole physics is the entropy of the system:
\begin{eqnarray}
S_\beta=\beta^2\partial_\beta
F_\beta=\frac{A_\perp}{90\beta\epsilon^2} \left
[\left(\frac{2\pi}{\beta}\right)^2+5\right].
\label{em32}
\end{eqnarray}
This equation gives, in Rindler space approximation, the one-loop
quantum correction to the black hole entropy coming form the
electromagnetic field propagating in the region outside the horizon.
It shows the well known horizon divergence \cite{TH} (see also 
\cite{BM} for a recent review on this topic): unless we
suppose the existence of a natural effective cutoff at the Planck scale
due to an (unknown) quantum gravity theory or back-reaction horizon
fluctuations etc.,\footnote{However, such a cutoff should depend on the
field spin value to produce the correct entropy factor in front of the
horizon area. See \cite{YO}.} we get a divergent entropy which is
physically unsatisfactory and contrasts with the finite
thermodynamical Bekenstein-Hawking entropy. However, this problem is
not peculiar to the photon field, as it occurs for scalar and
spinorial fields as well.

We can note that, if we took into account the surface term which we
have previously dropped,  we would obtain the unphysical, because being
gauge dependent, expression
\begin{eqnarray}
S_\beta(\alpha)=\beta^2\partial_\beta F_\beta=\frac{A_\perp}{90\beta\epsilon^2}
\left [\left(\frac{2\pi}{\beta}\right)^2+5\right] - (1-\frac{1}{2}\ln
\alpha)\frac{A_\perp}{6\beta\epsilon^2}.
\nonumber
\end{eqnarray}
As anticipated above, this expression for the entropy is negative when
the singularity is absent, $\beta = 2\pi$, and $\ln\alpha <
\frac{6}{5}$. Moreover, for $\ln\alpha<\frac{4}{3}$, $S_\beta(\alpha)$
shows a further zero of the entropy  corresponding to an inconsistent
(gauge-depending) {\em finite} temperature  {\em pure} quantum state
of the field.

Another thermodynamical quantity that we can compute from the
effective action (\ref{em31}) is the internal energy. Since it is well
known \cite{BD} that the usual Minkowski vacuum state, restricted to
the Rindler wedge, may be viewed as a Rindler  thermal state at
temperature $T=\frac{1}{2\pi}$, it is natural to require that the
internal energy vanishes when $\beta=2\pi$, namely, when the conical
singularity is absent. Hence, we define a renormalized free energy as
$F_\beta^{\mbox{\scriptsize sub.}}=F_\beta-U_{2\pi}$ which, by means
of the relation $U_\beta=\frac{1}{\beta}S_\beta+F_\beta$,
automatically gives $U_\beta^{\mbox{\scriptsize
sub.}}=U_\beta-U_{2\pi}$, that trivially vanish at $\beta=2\pi$, while
$S_\beta^{\mbox{\scriptsize sub.}}=S_\beta$. Explicitly,
\begin{eqnarray}
U_\beta^{\mbox{\scriptsize sub.}}=
\frac{\pi^2A_\perp}{30\beta^4\epsilon^2}+\frac{A_\perp}{36\beta^2\epsilon^2}-
\frac{13A_\perp}{1440\pi^2\epsilon^2}.
\end{eqnarray}
{From} this expression we can also compute the thermal
energy-momentum tensor: using the relation $U_\beta=-\int <T_0^0>
rdr\,dy\,dz$, supposing that $<T_0^0>$ depends on $r$
only\footnote{The remaining coordinates define Killing vectors.} and
that it vanishes at $\beta=2\pi$, we get
\begin{eqnarray}
<T_0^0>^{\mbox{\scriptsize sub.}}&=&-\frac{\pi^2}{15\beta^4 r^4}-
\frac{1}{18\beta^2 r^4}
+\frac{13}{720\pi^2 r^4},\nonumber\\
<T_{ab}>^{\mbox{\scriptsize sub.}}&=&
\frac{1}{3}<T_0^0>^{\mbox{\scriptsize sub.}}
[4\frac{{\cal K}_a{\cal K}_b}{{\cal K}^2}-g_{ab}],
\label{stress}
\end{eqnarray}
where in the last equation we have supposed a perfect fluid form,
${\cal K}_a=(\partial_t)_a$ is the time-like Killing vector associated
with the time coordinate of the Rindler space and ${\cal K}^2={\cal
K}_a{\cal K}^a$. This result for $<T_0^0>^{\mbox{\scriptsize sub.}}$ is
in agreement with twice the local heat kernel result \cite{CKV}.

As we have already said in the introduction, our results for the
thermodynamical quantities differ from those obtained with the point
splitting and the optical metric methods \cite{DC,BO,DO94,FS,ALO}. In
fact, for $<T_0^0>^{\mbox{\scriptsize sub.}}$ they give
\begin{equation}
-\frac{\pi^2}{15\beta^4r^4}-\frac{1}{6\beta^2
r^4}+\frac{11}{240\pi^2r^4},
\label{sp1}
\end{equation}
for spin $1$ and one-half this quantity for spin $0$. Our result for
the coefficient of the term proportional to $\beta^{-2}$ is one third
of that in Eq. (\ref{sp1}), while the difference in the numerical
coefficient of the term independent of $\beta$ is unimportant, since
it is determined by the others two by requiring the vanishing of the
energy-momentum tensor for $\beta=2\pi$. The reason of this
discrepancy, which appears also in the heat kernel approach
\cite{CKV,DK,LW,BS,FM} is not clear to us and requires further
investigations.

\section{A GENERAL CONJECTURE}

Let us focus our attention back on Kabat's surface term in the
effective lagrangian, Eq. (\ref{em30}): is it an accident which
appears in our manifold and in the vector case only, or conversely,
is it a more general phenomenon?

We can grasp some insight by studying either  the local
$\zeta$ function, as it appears in Eq. (\ref{emZETA}), or the local
heat kernel of Eq. (\ref{em12}) and passing to the local
zeta-functions through Eq. (\ref{em16}). In fact the Kabat term
already  comes out in the heat kernel and then it remains
substantially unchanged passing to the local $\zeta$ function through
Eq. (\ref{em16}).   The components of the modes $II$, $III$ and $IV$
contain (covariant) derivatives in both the conical and ${\mbox R}^2$
indices. Using trivial (covariant) derivative rules and reminding that
$\nabla_\mu \nabla^\mu \phi = -\lambda^2 \phi$ and $\partial_i
\partial^i \phi = -{\bf k}^2 \phi$ we may transform scalar products of
(covariant) derivatives  appearing in the integrand of Eq.
(\ref{em12}) into a  covariant divergence of a vector plus a simple
scalar term. Summing over the modes, these parts produce respectively
the Kabat surface term and the `twice scalar' part of the effective
Lagrangian in Eq. (\ref{em30})  (the mode $I$ gives a contribution to
this latter part only). This is the general mechanism which produces
Kabat's term. Let us  illustrate this in more detail. Dealing with the
modes $IV$ we find
\begin{eqnarray}
g^{ab} A^{(IV)\ast}_a A^{(IV)}_b
&=& \frac{1}{\lambda^2+{\bf k}^2}
\nabla_a \phi^\ast \nabla^a \phi \nonumber \\
&=&\frac{1}{\lambda^2+{\bf k}^2} \left[  \nabla_a (\phi^\ast
\nabla^a\phi ) - \phi^\ast \nabla_a \nabla^a \phi
\right] \nonumber\\
&=&\frac{1}{\lambda^2+{\bf k}^2} \left[ \nabla_a
(\phi^\ast \nabla^a \phi )  + (\lambda^2+ {\bf k}^2) \phi^\ast  \phi
\right]\:. \label{ema1}
\end{eqnarray}
Thus, using the particular form of our modes we get
\begin{eqnarray}
g^{ab} A^{(IV)\ast}_a A^{(IV)}_b
&=&\frac{1}{2(\lambda^2+{\bf k}^2)}\nonumber
\Delta
J_{\nu_n}^2 +  J^2_{\nu_n}  \:. \nonumber
\end{eqnarray}
The modes $III$  contribute to the local heat kernel and to the
effective Lagrangian in the same way. The modes $II$ require a  little
different care: we have
\begin{eqnarray}
g^{ab} A^{(II)\ast}_a A^{(II)}_b
&=&\frac{1}{\lambda^2} g^{\mu\nu}\epsilon_{\mu\sigma}
\epsilon_{\nu\rho} \nabla^\sigma \phi^\ast \nabla^\rho \phi \nonumber\\
&=&\frac{1}{\lambda^2} \left[  \nabla^\sigma ( g^{\mu\nu}
\epsilon_{\mu\sigma} \epsilon_{\nu\rho} \phi^\ast \nabla^\rho \phi
) - g^{\mu\nu}\epsilon_{\nu\rho} \epsilon_{\mu\sigma} \phi^\ast
\nabla^\sigma\nabla^\rho \phi \right] \nonumber\\
&=&\frac{1}{\lambda^2} \left[  \nabla^\sigma ( g_{\sigma\rho}
\phi^\ast \nabla^\rho \phi ) -\phi^\ast  g_{\rho\sigma} \nabla^\rho
\nabla^\sigma \phi \right] \nonumber\\
&=&\frac{1}{\lambda^2} \left[  \nabla_\mu (\phi^\ast
\nabla^\mu \phi)  + \lambda^2 \phi^\ast  \phi \right]\nonumber\\
&=&\frac{1}{\lambda^2} \left[  \nabla_a (\phi^\ast
\nabla^a \phi)  + \lambda^2 \phi^\ast  \phi \right]\:. 
\label{ema2}
\end{eqnarray}
And thus, reminding the particular form of our modes
\begin{eqnarray}
g^{ab} A^{(II)\ast}_a A^{(II)}_b
&=&\frac{1}{2\lambda^2} \nonumber
\Delta
J_{\nu_n}^2 +  J^2_{\nu_n} \nonumber \:.
\end{eqnarray}
The contribution to the effective Lagrangian is similar to the
previous ones. In both the examined cases, using the specific form of
scalar eigenfunctions, we have obtained the right-hand side of
Eq. (\ref{em25}) except for some factors  which will be arranged
summing over all the modes in the final result.
The term  $\nabla_a (\phi^\ast \nabla^a \phi)$ $(= \frac{1}{2}\Delta
J^2_{\nu_n})$ contributes only to the second term of the right-hand
side of  Eq. (\ref{ema0}), namely it contributes only to the Kabat
surface term in the effective Lagrangian in Eq. (\ref{em30}). Moreover,
the  term $\lambda^2 \phi^\ast \phi$ $(= \lambda^2 J^2_{\nu_n})$
contributes only to the remaining term in the right hand side of
Eq. (\ref{ema0}) and thus to the  ``twice scalar'' part of the same
effective Lagrangian only.

We  further remark that the previously employed  covariant derivative
identities are exactly the same which one has to use in order to check
the correct normalization of the modes.\footnote{In this case the
indices $(n\lambda {\bf k})$ which appear in the modes $A_a$ and
$A^\ast_a$ are generally different.} However, in that case the
surface terms are dropped after the formal integration in the  spatial
variables, because they do not contribute, in a distributional sense,
to the overall normalization. Conversely, following the local zeta
function method they produce Kabat-like terms.

More generally speaking, following the previous outline, one
can avoid specifying the form of the scalar eigenfunction and the use
of Eq. (\ref{em25}), remaining on a more general ground.\footnote{It is
clear from our discussion that the Kabat term gets contributions from
each mode $II,III,IV$ not depending on the corresponding eigenvalue.
This term does not coincide  with the surface term recently suggested
by  Fursaev and  Miele \cite{FM} dealing with  compact
manifolds, because this latter involves zero modes only.} This means
that we can consider a more general manifold which is topologically
${\cal M}\times \mbox{R}^2$ with the natural product metric, where
${\cal M}$ is any, maybe curved, two-dimensional manifold. The photon
effective action can be written  as
\begin{eqnarray}
\ln Z = -\frac{1}{2} \ln \det \mu^{-2} \left(
-\Delta_1 + (1- \frac{1}{\alpha}) d_0 \delta_0
\right) + \ln Z_{\mbox{\scriptsize ghost}}
\label{ema3}\:,
\end{eqnarray}
where $\Delta_1 = d_0\delta_0 +\delta_1 d_1$ is the  Hodge Laplacian
for 1-forms ($\delta_n \equiv d_n^\dagger $ with respect to the Hodge
scalar product.)
The eigenfunctions of the operator appearing in the above equation can
still be written as in Eq. (\ref{em20b}). Now, $\phi = \frac{1}{2\pi}
e^{ik_y y+ i k_z z} {\bf J}_{n,\lambda}(x^\mu)$ where ${\bf
J}_{n,\lambda}(x^\mu)$ is an eigenfunction of (the Friedrichs
extension of) the 0-forms Hodge Laplacian\footnote{Remind the Hodge
Laplacian coincides with minus the Laplace-Beltrami operator for
0-forms. This generally  does not happen for n-forms when $n>0$ in
curved manifolds.}
$\Delta^{\cal M}_0$ on ${\cal M}$, with eigenvalue $-\lambda^2$.
Employing a bit of $n$-forms algebra, one can obtain in our manifold
the same eigenvalues found in the manifold ${\cal C}_\beta \times
\mbox{R}^2$. Furthermore, once again $\delta_0 A^{(y)} = 0$,  namely $
\nabla^a A_a^{(y)}= 0 $, in case $y= I, II, III$. Then, using Eq.s
(\ref{ema1}) and (\ref{ema2}) and the definition in Eq.
(\ref{emZETA}), we get, before we take into account the ghosts
contribution,
\begin{eqnarray}
\zeta_{{\cal M}\times \mbox{\scriptsize R}^2}(s;x) = (3+\alpha^s)
\zeta^{\mbox{\scriptsize scalar}}_{{\cal M}\times \mbox{\scriptsize
R}^2}(s;x) + \frac{s+1+\alpha^s(s-1)}{2s} \zeta^{V}_{{\cal M}\times
\mbox{\scriptsize R}^2}(s;x) \:, \nonumber
\end{eqnarray}
where the surface term reads
\begin{eqnarray}
\zeta^{V}_{{\cal M}\times \mbox{\scriptsize R}^2}(s;x)
=\frac{\Gamma(s-1)}{4\pi \Gamma(s)} \nabla_a \sum_n \int d\lambda
\lambda \,{\bf J}^\ast \nabla^a {\bf J}\:.\nonumber
\end{eqnarray}
Notice that, if the manifold is regular and compact, this surface term
automatically disappears after we integrate over the spatial
variables. Instead, if the manifold ${\cal M}$ has conical
singularities or boundaries, then this term could survive the
integration.  We can further  suppose  that ${\cal M}$ contains a
Killing vector $\partial_\tau$ with compact orbits in such a manner
that we can define a temperature $1/\beta$ and interpret the effective
action as the logarithm of the photon partition function. Employing
coordinates $r, \tau$ on ${\cal M}$, we can decompose ${\bf
J}_{n,\lambda}(r,\tau)$ as ${\bf J}_{n,\lambda}(r,\tau) =
{\beta}^{-1/2}e^{-2\pi n i\tau/\beta} {\cal J}_{n,\lambda}(r)
$, ${\cal J}_{n,\lambda}(r)$ being real. The surface term reads, in
this case,
\begin{eqnarray}
\zeta^{V}_{{\cal M}\times \mbox{\scriptsize R}^2}(s;x)
=\frac{\Gamma(s-1)}{4\pi \beta \Gamma(s)} \Delta_0 \sum_n
\int d\lambda \lambda\, {\cal J}_{n,\lambda}(r)^2 \:. \nonumber
\end{eqnarray}
Equation (\ref{ema3}) holds in very general manifolds, also dropping the
requirement of a metric which is Cartesian product of the flat
$\mbox{R}^2$ metric and any other metric.

One can simply prove that, if $\phi$ is an eigenfunction of $\Delta_0$
with eigenvalue $-\nu^2$ on such a general manifold, $A = d_0 \phi$
will be an eigenfunction of the vector operator $-\Delta_1 +
(1-\frac{1}{\alpha})d_0\delta_0$ with gauge-dependent eigenvalue
$-\nu^2/\alpha$. Employing the rule in Eq. (\ref{ema1}) with $\nu^2$
in place of $\lambda^2 + {\bf k}^2$, we expect that this latter
eigenfunction should produce a (gauge-dependent) surface term into the
local zeta function.

Dealing with spin $s\geq 1$ and massless fields, because of  the
simple equation of motion form (in Feynman-like gauges at least), we
expect to find out some normal modes obtained as covariant derivatives
of the scalar field modes opportunely rearranged. Hence, barring
miraculous cancellations, the corresponding local heat kernel, local
$\zeta$ function and effective Lagrangian, should contain Kabat-like
surface terms, due to the previous mechanism. We will check this for
the graviton in the next section.\footnote{We also tried  to study the
photon case employing  a so-called  `physical gauge' as $A_z=0$. The
use of the $\zeta$-function regularization in this case is problematic
due to a remaining gauge ambiguity arising  whenever one tries to deal
with a path integral nonformal approach in axial gauges.
Nevertheless, through the same mechanism, the Kabat term seems to
survive  in this case as well. }

\section{THE GRAVITON $\zeta$ FUNCTION IN THE HARMONIC GAUGE}

In this section we shall compute the local $\zeta$ function in the case
of a linearized graviton propagating in the Rindler wedge. We will see
that Kabat-like surface terms indeed appear, as we suggested in
the previous section. Moreover, we will find out that consistent
results arise by discarding all those terms.

Following the same procedure  used in \cite{HA,GI}, which employs the
harmonic gauge, we decompose the linearized field of a graviton into
its symmetric traceless part $h_{ab}$ and its trace part $h$. Choosing
an opportune normalization factor of the fields and dropping boundary
terms, the Euclidean action (containing also the gauge-fixing part)
looks such as:
\begin{eqnarray}
S_E[h_{ab},h] = \frac{1}{32 \pi G}\int dx^4 \sqrt{g}\left\{
\frac{1}{2} g^{aa'}g^{bb'} h_{ab}\nabla_c \nabla^c
h_{a'b'} +\frac{1}{4} h \nabla_d\nabla^d h\right\},
\label{lagranggrav}
\end{eqnarray}
where $g$, $g^{ab}$, and covariant derivatives are referred to the
background metric, namely, the Euclidean Rindler metric. That metric is
also used to raise and lower indices. Notice that curvature tensor
terms (see \cite{GI}) do not appear in the above action and this is
due to the flatness of the manifold. It is necessary to point out that
we changed the sign of  the trace field Lagrangian as this  appeared
after we performed a ``simple'' Wick rotation toward the imaginary time
on the Lorentzian Lagrangian. In fact, in order to obtain an Euclidean
Lagrangian producing a formally finite functional
integral\footnote{Remind that this functional integral contains the
exponential $\exp {(- S_E)}$}, it is also necessary to rotate the
scalar field $h$  into imaginary values during the Wick rotation. This
adjusts the sign in front of the corresponding Lagrangian
\cite{HA,GI}. We can write, as far as the effective action is
concerned:
\begin{eqnarray}
\ln Z_{\mbox{\scriptsize gravitons}}
= -\frac{1}{2}\ln \det \mu^{-2} \left[ -g^{aa'}g^{bb'} \nabla_c \nabla^c
\right]-\frac{1}{2}\ln \det \mu^{-2}\left[ -\nabla_d \nabla^d
\right] + \ln Z_{\mbox{\scriptsize grv. ghosts}}
\label{zetagrav}\:.
\end{eqnarray}
The first determinant has to be evaluated in the $L^2$ space of
traceless symmetric tensorial field. Unessential factors in front of
the operators can be dropped into an overall added constant and thus
omitted. Furthermore, the ghost contribution has been taken into
account through the last term of the previous equation. A usual
procedure\footnote{This result holds also for local quantities.}
leads us to \cite{HA,GI}
\begin{eqnarray}
\ln Z_{\mbox{\scriptsize grv. ghosts}}
= - 2 \ln Z_{\mbox{\scriptsize vector}}.
\nonumber
\end{eqnarray}
The  partition function in  $\ln Z_{\mbox{\scriptsize vector}}$ is the
partition function obtained quantizing the massless Klein-Gordon
vector field. Hence,  this also coincides with the photon partition
function evaluated in the Feynman gauge, namely $\alpha=1$ in Eq.
(\ref{em10}), {\em without} taking into account the photon ghost
contribution. Thus, from the effective graviton ghost action, two
vector $\alpha=1$ Kabat's surface terms (with the sign changed) arise.
In order to compute the above functional determinants, we have to look
for normalized modes of a self-adjoint extension of the  tensorial
Laplace-Beltrami operator $\Delta_T = g^{aa'}g^{bb'} \nabla_c
\nabla^c$ in the space of symmetric traceless tensors  and the scalar
Laplace-Beltrami operator $\Delta_S = \nabla_d \nabla^d$. Obviously,
the eigenfunctions of $\Delta_S$ can be chosen as $h_{n\lambda {\bf
k}} = \phi_{n\lambda {\bf k}}(x) $, where, as before, $\phi =
\phi_{n\lambda k}(x)$ indicates the generic eigenfunction of the
scalar Laplacian, Eq. (\ref{em21}).

In the tensorial case, we find the following nine classes of
sym\-metric trace\-less eigen\-functions:\footnote{All the components
of each eigenfunction class which do not appear in the following list
are understood to vanish.}
\begin{eqnarray}
h^{(1)}_{n\lambda {\bf k}}&:&
\:\:\:\: \frac{\sqrt{2}}{\lambda^2}\nabla_\mu \nabla_\nu
\phi + \frac{1}{\sqrt{2}} g_{\mu\nu}
\phi = h^{(1)}_{\mu\nu}=h^{(1)}_{\nu\mu}\:;\nonumber \\
h^{(2)}_{n\lambda {\bf k}} &:&
\:\:\:\: \frac{\sqrt{g}}{\sqrt{2}\lambda^2}\left\{
\epsilon_{\mu\sigma}\nabla^\sigma \nabla_\nu\phi +
\epsilon_{\nu\sigma}\nabla^\sigma \nabla_\mu\phi \right\}
= h^{(2)}_{\mu\nu}=h^{(2)}_{\nu\mu}\:;\nonumber \\
h^{(3)}_{n\lambda {\bf k}}&:&\:\:\:\:
\frac{1}{\sqrt{2}k\lambda} \partial_i \nabla_\mu \phi = h^{(3)}_{i\mu}
= h^{(3)}_{\mu i} \:;\nonumber \\
h^{(4)}_{n\lambda {\bf k}}&:& \:\:\:\:
\frac{\sqrt{g}}{\sqrt{2} k\lambda}  \epsilon_{\mu\nu}\partial_i
\nabla^\nu \phi = h^{(4)}_{i\mu}= h^{(4)}_{\mu i} \:;\nonumber \\
h^{(5)}_{n\lambda {\bf k}} &:& \:\:\:\:
\frac{\sqrt{g}}{\sqrt{2} k\lambda}  \epsilon_{\mu\nu}\epsilon_{ij}
\partial^j \nabla^\nu\phi = h^{(5)}_{i\mu}
= h^{(5)}_{\mu i} \:;\nonumber \\
h^{(6)}_{n\lambda {\bf k}} &:&\:\:\:\:
\frac{1}{\sqrt{2} k\lambda} \epsilon_{ij}
\partial^j \nabla_\mu\phi = h^{(6)}_{i\mu}
= h^{(6)}_{\mu i} \:;\nonumber \\
h^{(7)}_{n\lambda {\bf k}} &:&
\:\:\:\: \frac{\sqrt{2}}{ {\bf k}^2}\partial_i \partial_j
\phi + \frac{1}{\sqrt{2}} \delta_{ij} \phi =
h^{(7)}_{ij}=h^{(7)}_{ji}\:;\nonumber \\
h^{(8)}_{n\lambda {\bf k}} &:&
\:\:\:\: \frac{1}{\sqrt{2}{\bf k}^2}\left\{
\epsilon_{ik}\partial^k\partial_j\phi +\epsilon_{jk}
\partial^k\partial_i\phi\right\} = h^{(8)}_{ij}=h^{(8)}_{ji}\:;\nonumber \\
h^{(9)}_{n\lambda {\bf k}} &:&
\:\:\:\: \frac{1}{2} g_{\mu\nu}\phi -\frac{1}{2}\delta_{ij}\phi =
h^{(9)}_{ab}
\:.\nonumber
\end{eqnarray}
Here, $\sqrt{g}\epsilon_{\mu\nu}$ indicates the antisymmetric
Levi-Civita pseudotensor on the cone and $\epsilon_{ij}$ the
antisymmetric Levi-Civita pseudotensor on ${\mbox{R}}^2$ in Cartesian
coordinates. The previous modes satisfy
\begin{eqnarray}
\Delta_T h^{(y)}_{n\lambda {\bf k}} = -(\lambda^2 + 
{\bf k}^2) h^{(y)}_{n\lambda {\bf k}},
\hspace{1cm} y= 1,2,... ,9,
\end{eqnarray}
and
\begin{eqnarray}
\Delta_S h_{n\lambda {\bf k}} = -(\lambda^2 + {\bf k}^2)
h_{n\lambda {\bf k}} \:.
\end{eqnarray}
Finally, the normalization relations are ($ y,y'= 1,2....,9$)
\begin{eqnarray}
\int d^4x \sqrt{g} \: g^{aa'}g^{bb'} \: h^{(y)\ast}_{n\lambda {\bf k}}
(x)_{ab}
h^{(y')}_{n'\lambda' k'_t}(x)_{a'b'} =
\delta^{yy'}\delta_{nn'}
\delta^{(2)}({\bf k}- {\bf k}')
\frac{\delta (\lambda-\lambda')}{\lambda}\nonumber
\end{eqnarray}
and
\begin{eqnarray}
\int d^4x \sqrt{g} h^{\ast}_{n\lambda {\bf k}}(x)
h_{n'\lambda' k'}(x) = \delta_{nn'}
\delta^{(2)}({\bf k}- {\bf k}')
\frac{1}{\lambda}
\delta (\lambda-\lambda')\:. \nonumber
\end{eqnarray}
Using Eq. (\ref{emZETA}), we can write the local $\zeta$ function as
\begin{eqnarray}
\zeta^{\mbox{\scriptsize Gravitons}}(s;x)&=& \sum_{y=1}^9
\zeta^{(y)}(s;x) + \zeta^{\mbox{\scriptsize scalar}}(s;x)\nonumber\\
&=&\sum_{y=1}^9
\sum_{n=-\infty}^{\infty}\int_0^{\infty}d\lambda\,\lambda
\int_{\mbox{R}^2} d^2{\bf k}\, \nu_{n}^{-2s}
g^{aa'}(x) g^{bb'}(x) h^{\ast(y)}(x)_{ab}\,h^{(y)}(x)_{a'b'}+\nonumber\\
& &+ \sum_{n=-\infty}^{\infty}\int_0^{\infty}d\lambda\,\lambda
\int_{\mbox{R}^2} d^2{\bf k}\,\nu_{n}^{-2s} h^\ast(x) h(x)
\label{gravheat}.
\end{eqnarray}
The latter term takes into account the graviton trace part
contribution to local $\zeta$ function. Obviously, this is exactly the
scalar local $\zeta$ function. Let us rather  consider the former term
and, in particular, the contribution due to $h^{(1)}$. Following the
sketch of the previous section, we can rearrange this term
transforming the product of the covariant derivatives into a scalar
term added to several covariant divergences of vector and tensor
fields:
\begin{eqnarray}
\zeta^{(1)}(t;x)&=& \frac{\Gamma(s-1)}{4\pi \beta \Gamma(s)}
\sum_n
\int_0^{\infty}d\lambda\,\lambda^{3-2s} \,
\phi^{\ast} \phi\nonumber\\
& &+ 4 \frac{\Gamma(s-1)}{4\pi \beta \Gamma(s)}\sum_n
\int_0^{\infty}d\lambda\,\lambda^{1-2s}\,
\nabla_a (\phi^\ast \nabla^a \phi)+ \nonumber\\
& &+ 2 \frac{\Gamma(s-1)}{4\pi \beta \Gamma(s)}
\sum_n
\int_0^{\infty}d\lambda\,\lambda^{-2s} \,
\nabla^a \nabla_b [\nabla_a \phi^\ast
\nabla^b \phi]\:.\nonumber
\end{eqnarray}
Using different notation, we finally find
\begin{eqnarray}
\zeta^{(1)}(s;x)= \zeta^{\mbox{\scriptsize scalar}}(s;x)+ 2
\zeta^{V}(s;x)+  2 \zeta^{W}(s;x)
\end{eqnarray}
where we defined
\begin{eqnarray}
\zeta^{W}(s;x) &=& \frac{\Gamma(s-1)}{4\pi \beta \Gamma(s)} \sum_n
\int_0^{\infty}d\lambda\,\lambda^{-2s} \, \nabla^a \nabla_b
(\nabla_a \phi^\ast \nabla^b \phi )\nonumber\\
&=&\frac{\Gamma(s-1)}{4\pi \beta \Gamma(s)} \sum_n
\int_0^{\infty}d\lambda\,\lambda^{-2s} \nonumber\\
& &\times\left[ \frac{1}{r}\partial_{r} r \partial_{r} (\partial_r
J_{\nu_n})^2 + \frac{1}{r}\partial_{r}(\partial_r J_{\nu_n})^2 -
\frac{\nu_n^2}{r} \partial_r \frac{J_{\nu_n}(\lambda r)^2}{r^2}\right]
\:.
\end{eqnarray}
Thus, we see that in the local $\zeta$ function the $(\alpha=1)$-Kabat
surface term $\zeta^{V}(s;x)$ reappears, together with a new surface
term $\zeta^{W}(s;x)$. The contribution of $h^{(2)}$ is similar to
the previous one  and it reads
\begin{eqnarray}
\zeta^{2}(s;x)=
\zeta^{\mbox{\scriptsize scalar}}(s;x)+ \zeta^{V}(s;x)+\zeta^{W}(s;x)+
\zeta^{U}(s;x) \nonumber
\end{eqnarray}
where, provided $\theta^{ab} =\epsilon^{ab}$ when $a,b = \mu, \nu $
and $\theta^{ab} = 0 $ otherwise,
\begin{eqnarray}
\zeta^{U}(s;x)&=&
\frac{\Gamma(s-1)}{4\pi \beta \Gamma(s)} \sum_n
\int_0^{\infty}d\lambda\,\lambda^{-2s} \, \nabla_a \nabla_b \left[
g\:\theta^{ac}\theta^{bd} \nabla_c \phi^\ast
\nabla_d \phi     \right]\nonumber \\
&=& \frac{\Gamma(s-1)}{4\pi \beta \Gamma(s)}
\sum_n
\int_0^{\infty}d\lambda\,\lambda^{-2s} \,
\partial_r \left[
\frac{J_{\nu_n}^2}{r}-(\partial_r J_{\nu_n})^2 \right]. \nonumber
\end{eqnarray}
The contributions of the remaining terms are much more trivial.
In fact, a few of algebra leads us to
\begin{eqnarray}
\zeta^{(3)}(s;x)&
=&\zeta^{(4)}(s;x)=\zeta^{(5)}(s;x)=\zeta^{(6)}(s;x)\nonumber\\
&=& \zeta^{\mbox{\scriptsize scalar}}(s;x)+ \frac{1}{2} \zeta^{V}(s;x),
 \nonumber
\end{eqnarray}
and
\begin{eqnarray}
\zeta^{(7)}(s;x)=\zeta^{(8)}(s;x)=\zeta^{(9)}(s;x)=
\zeta^{\mbox{\scriptsize scalar}}(s;x).
\nonumber
\end{eqnarray}
Finally, we have already noted above that the contribution of the trace
terms $h$ is
exactly $\zeta^{\mbox{\scriptsize scalar}}(s;x)$.
Then, taking into account the contribution of the
ghost Lagrangian, which amounts
to $-8 \zeta^{\mbox{\scriptsize scalar}}(s;x) -
2\zeta^{V}(s;x) $, we
get the final expression of spin-traced graviton $\zeta$ function:
\begin{eqnarray}
\zeta^{\mbox{\scriptsize Gravitons}}(s;x)=
2 \zeta^{\mbox{\scriptsize scalar}}(s;x)+
3 \zeta^{V}(s;x)+
3 \zeta^{W}(s;x) +\zeta^{U}(s;x).
\end{eqnarray}
Dropping the last three surface terms we obtain the reasonable
result which agrees with the counting of the true graviton degrees of
freedom: ${\cal L}^{\mbox{\scriptsize Graviton}}(x) = 2 {\cal L}^{
\mbox{\scriptsize scalar}}(x)$. Hence, all the thermodynamical
quantities coincides with those of the previously computed photon
fields.

\newpage

\section{DISCUSSION}

In this paper we have computed the effective action of the photon and
graviton fields in the conical background $C_\beta\times\mbox{R}^2$,
and our main result is that it is just what one expects from counting
the number of degrees of freedom, i.e. twice that of the massless
scalar effective action. Moreover, we have got the correct Planckian
temperature dependence of the thermodynamical quantities.

To get this apparently trivial result, we had to deal with unwanted
terms arising from the presence of the conical singularity. We
discussed how the appearance of those surface terms is quite a general
phenomenon dealing with general manifolds in the case of fields with
integer non-zero spin. The presence of conical singularities needs
some further regularization procedure. In particular, this is
necessary while studying the photon field in order to restore the
gauge invariance of the integrated quantities. It could be interesting
to develop an analogue research in the case of gravitons in any
covariant gauge.

In the general case our proposal is the simplest one, namely, to
discard all the surface terms. However, we think that, away from our
local $\zeta$-function  approach,  this should not be the only possible
treatment of surface terms. In fact, comparing our results with
Kabat's it arises that, except for the two-dimensional case,  the
necessary treatment of surface terms strongly depends on the general
approach used to define and calculate the effective Lagrangian.
Moreover, it also depends on the regularization procedure used to
define the integrated quantities.

In our local $\zeta$-function approach, the meaning of the only cutoff
as the minimal distance from the horizon leads ourselves towards the
simple procedure of discarding the surface terms in order to restore
the gauge invariance. In Kabat's treatment, the meaning of the
employed cutoff is not so strict and permits one to make safe the
gauge invariance  and take on the surface terms as well. This is due
to fine-tuning of mode depending cutoffs which contain a further 
gauge-fixing parameter dependence.

In our approach, when integrating the surface term it is not possible
to use an $\alpha$-dependent cutoff different from that used for
the rest and such that it cancels the $\alpha$ dependence in the
integrated quantity: in fact, no real function  $\epsilon(\alpha)$ can
absorb the factor $(1-\frac{1}{2}\ln \alpha)$, appearing in the
integrated surface term, for all the values of $\alpha$.

In any case, we think that any procedure which does not discard the
surface terms must be able to explain why the consequent result is not
in agreement with what one expects from counting the number of degrees
of freedom and  to deal with the apparently unphysical corrections to
the thermodynamical quantities arising from those terms. Maybe this is
possible in an effective low-energy string theory which does not
coincide with the ordinary quantum field theory.

\par \section*{Acknowledgments}
We are grateful to  Guido Cognola,
Giuseppe Nardelli, and, in particular, to Luciano Vanzo  and  Sergio
Zerbini for valuable discussions and useful suggestions.

\end{document}